# Control of Radiation Damage in MoS$_2$ by Graphene Encapsulation

*Recep Zan[1,2,\*], Quentin M. Ramasse[3,\*], Rashid Jalil[1], Thanasis Georgiou[1], Ursel Bangert[2]*

*and Konstantin S. Novoselov[1,\*]*

[1]School of Physics and Astronomy, University of Manchester, Manchester, M13 9PL, U.K.

[2]School of Materials, University of Manchester, Manchester, M13 9PL, U.K.

[3]SuperSTEM Laboratory, STFC Daresbury Campus, Daresbury WA4 4AD, U.K.

*To whom correspondence should be addressed: recep.zan@manchester.ac.uk,

qmramasse@superstem.org and kostya@manchester.ac.uk

**Abstract**

Recent dramatic progress in studying various two-dimensional (2D) atomic crystals and their heterostructures calls for better and more detailed understanding of their crystallography, reconstruction, stacking order, *etc*. For this, direct imaging and identification of each and every atom is essential. Transmission Electron Microscopy (TEM) and Scanning Transmission Electron Microscopy (STEM) are ideal, and perhaps the only tools for such studies. However, the electron beam can in some cases induce dramatic structure changes and radiation damage becomes an obstacle in obtaining the desired information in imaging and chemical analysis in the (S)TEM. This is the case of 2D materials such as molybdenum disulfide MoS$_2$, but also of many biological specimens, molecules and proteins. Thus, minimizing damage to the specimen is essential for optimum microscopic analysis. In this letter we demonstrate, on the example of MoS$_2$, that encapsulation of such crystals between two layers of graphene allows for a dramatic improvement in stability of the studied 2D crystal, and permits careful control over the defect nature and formation in it. We present STEM data collected from single layer MoS$_2$ samples prepared for observation in the microscope through three distinct procedures. The fabricated single layer MoS$_2$ samples were either left bare (pristine), placed atop a single-layer of graphene or finally encapsulated between single graphene layers. Their behaviour under the electron beam is carefully compared and we show that the MoS$_2$ sample 'sandwiched' between the graphene layers has the highest durability and lowest defect formation rate compared to the other two samples, for very similar experimental conditions.

Similar to graphene, other 2D nanomaterials such as transition metal dichalcogenides (TMD) have attracted considerable attention due to their unique physical, chemical, and structural properties as well as their great potential for applications.[1-3] MoS$_2$ is one of the better known and most studied of this new class of layered TMD materials and is already widely used in industrial processes as a nano-catalyst and dry lubricant.[4,5] Owing to the unique properties of single layer MoS$_2$, and in particular the fact that it possesses a direct band gap,[6-8] much effort has been put into exploring its potential use in combination with graphene in electronic and optoelectronic devices with a view to build logic transistors, field emitters and detectors.[9] Single layer MoS$_2$ consists of two distinct sub-lattices: Mo (metal) atoms are trigonal-prismatically bonded to pairs of S atoms and arranged hexagonally in plane. The resulting S-Mo-S slabs are then stacked to various degrees, 2H stacking being the most common. Similar to graphite and h-BN, chemical bonds in-plane (within the slab) are covalent (strong) while weak van der Waals bonds are established between stacked layers to form a bulk (3D) crystal. As for graphene, this enables single MoS$_2$-layer mechanical exfoliation,[1] but there are several other fabrication methods such as liquid exfoliation [10] and chemical vapour deposition.[11,12]

In conductive materials such as graphene, 'knock-on' (the displacement of atoms from their original positions in the lattice) is the dominant radiation damage mechanism during observation in an electron microscope, whereas ionization damage (radiolysis) prevails in semiconductors and insulators.[13] If knock-on is the dominant damage mechanism, reducing the primary beam energy below the knock-on threshold will prevent it.[14,15] If radiolysis is the main damage process, there is no sharp energy threshold below which no damage occurs although cooling the specimen can help to reduce it. Radiation damage in single layer MoS$_2$ has so far not been investigated to a great extent despite the fact that being able to control the defect formation is crucial: the presence of a controlled number of defects in 2D materials can lead to physical properties entirely different from their pristine form.[16] Furthermore, if optimal beam-damage prevention is achieved, the time to acquire an image can be increased until a sufficient signal-to-noise (S/N) ratio is obtained.[13] *Ab initio* studies of the displacement threshold of S atoms in pristine MoS$_2$ layer report a knock-on energy of ~6.5 eV.[17, 18] Interestingly, this corresponds approximately to the maximum knock-on energy transfer from 80 keV electrons to $^{32}$S atoms while 60 keV electrons can transfer up to 4.3 eV to $^{32}$S (and 7.4 eV at 100 keV). In TEM observations of single layer MoS$_2$ both knock-on and ionization damage mechanisms are thus likely at play although at the low primary beam energies often used to image 2D materials the ionization is probably dominant.[19, 20] Atom by atom investigation is nonetheless difficult without any protective measure.

A commonly used method to preserve the crystalline form of a specimen over extended electron exposure times, and hence to avoid damage be-it caused by beam knock-on or through radiolysis, is to encapsulate the sample within a protective (conductive) layer.[21] This is typically achieved for TEM observation by evaporating thin layers of amorphous carbon. This technique has obvious drawbacks, chief of which is the reduced contrast from the actual specimen because of its carbon coating. For 2D materials, one to a few atoms thick, such an approach would be almost impossible. Instead, we demonstrate here the use of graphene to encapsulate the MoS$_2$ crystal. Graphene, which has also been suggested as a support for nano- and bio-particle imaging,[22, 23] and used as membrane for liquid cells for in-situ TEM studies,[24-26] generates extremely low background signals compared to amorphous carbon films, resulting in a dramatic increase in the S/N ratio in (S)TEM micrographs. Such an approach not only protects the MoS$_2$ from environmental effects such as chemical etching, but also, thanks to graphene's excellent electrical and thermal conductivity, increases the stability of the MoS$_2$ layer under the electron beam through minimisation of charging effects and vibrations. Furthermore, its crystalline structure can be easily subtracted from the (S)TEM micrographs by Fourier filtering, thus enabling the visualisation of the particles as if surrounded by vacuum.[27, 28]

**Results and Discussion**

Our initial investigations, carried out on the bare (pristine) single layer MoS$_2$ sample, revealed that even with a 60 keV beam and near-UHV conditions, defect formation to a free-standing sample was unavoidable. Ab-initio modelling predicts this beam energy is much lower than the calculated displacement threshold for S atoms in single layer MoS$_2$,[17, 18] and severe knock-on damage is thus unlikely in our experimental conditions. We therefore suggest that ionization damage, which has severe effects on semiconductors and insulators in comparison to knock-on damage, is responsible for the defect creation. Figure 1 illustrates how imaging at high magnification immediately initiated damage: the image in fig. 1b was acquired approximately 30 s after that in fig. 1a, demonstrating the ease with which damage is introduced. The dwell time per pixel was 39 µs and the pixel size 0.098 Å corresponding to an electron dose of 2.5x10$^6$ e/Å$^2$. Following the vacancy formation in fig. 1b, a movie was recorded by drastically reducing the scanning time per frame (5.1µs dwell time per pixel, 256x256 pixels only) to observe dynamically any further damage progression (see supplementary information movie_1). From the movie, it can be seen that as soon as the MoS$_2$ sheet is perforated by losing first a single S atom and then the other Mo-bonded S atom,[17] this hole enlarges easily under the scanning probe. Not only are atoms at the edge of the hole less coordinated, thus facilitating ionization damage, it is also likely that the knock-on threshold of edge atoms is significantly lowered compared to the bulk allowing this mechanism to also come into play. Due to the higher displacement threshold of Mo compared to S, and to the propensity of Mo atoms to form metallic clusters, Mo atoms appear to aggregate on the edges of the damaged area while S atoms are simply sputtered away (this suggestion is supported by EELS measurements, not shown here). These heavy, bright aggregates are clearly seen in fig. 1d and g.[18] The first and last frames were extracted from movie_1 and are shown in figs 1c and d, respectively. The images, taken 65 s apart, show how the initial vacancy expands in a short time to a 2nm hole even under a relatively low (in materials science terms) electron dose-rate (1x10$^5$ e/Å$^2$/frame). High dose exposures are often required in order to obtain good enough S/N ratios in chemical (spectral) analyses. It is therefore instructive to scrutinise the HAADF images before and after an attempt at acquiring an electron energy loss spectrum image (SI) as they are shown in fig. 1e and 1g, respectively. The map was collected on the blue-framed rectangular area in the fig. 1e (survey image), and the HAADF image acquired simultaneously with the SI is presented in fig. 1f. The Gatan Enfina spectrometer was configured to acquire one spectrum every 50ms, corresponding to a total dose for the map of 2.6x10$^{10}$ e/Å$^2$. Note that even though modern energy loss spectrometers allow for much shorter pixel dwell times (and therefore lower dose-rates), resulting maps often suffer from poor signal-to-noise (S/N) ratio and for 2D materials in particular 50ms/spectrum was found to be the minimum for interpretable signal. Even though no damage is visible in the region of the SI on the survey image, taken immediately before the start of the EELS acquisition, the 'after' image shows that a large hole has been created. The HAADF image taken simultaneously with the EELS map reveals that the hole opened up very early into the SI acquisition and the damage became very severe, making this dataset unexploitable for chemical analysis (the resulting chemical maps are shown in Supplementary material).

(figure 1 here)

By contrast, graphene is known to be remarkably stable under the electron beam in very similar experimental conditions. Thanks to the combination of low primary beam energy (60 keV) and clean vacuum at the sample (below $5 \times 10^{-9}$ Torr), clean patches of pristine single-layer graphene can thus be observed repeatedly at very high electron doses and without any observable damage formation, even for extremely long periods of time.[14, 29, 30] A possible way to minimize the damage in $MoS_2$ (or in other beam sensitive materials), whilst perhaps also increasing the stability of the flake against vibrations under the beam, is therefore to use graphene as support.[23] To verify the presence of both single layer graphene and $MoS_2$ in the samples fabricated, a diffraction pattern was obtained on this double layer stack (graphene/$MoS_2$); it is presented in fig. 2a and reveals two clear sets of diffraction spots. The $MoS_2$ diffraction spots are more intense than those of graphene, and they form the very inner circle (red dashed on fig. 2a) of the diffraction pattern due to the larger lattice constant of $MoS_2$ ($d_{100}$= 2.7 Å) compared to graphene. The HAADF images presented in Figs. 2b and 2c were taken in this region with a dose of $5.1 \times 10^6$ e/Å$^2$, slightly higher than the dose applied to the bare $MoS_2$ sample (and in otherwise identical conditions): remarkably, no damage is observed. $MoS_2$ placed on graphene thus appears more stable under the scanning electron beam and provides longer working time without defect formation compared to bare $MoS_2$. Damage occurred only after longer electron beam exposure during the acquisition of a movie (see movie_2 in supplementary information, acquired in similar conditions to those employed for movie_1): figures 2d and 2e are the first and last frames of movie_2, taken 243s apart. The movie was taken with ~$3.8 \times 10^5$ e/Å$^2$ dose per frame (again slightly higher than the one used in movie_1), but at identical frame rate. The hole formation took longer than in the pristine sample even at higher applied dose (the first sign of damage was seen at frame 56, corresponding to a total accumulated dose of $2.3 \times 10^8$ e/Å$^2$), but once the damage started in the MoS$_2$, the expansion of the defective region was again quite rapid. Note the presence in fig. 2e of bright Mo atoms in the middle of the 'hole' confirming that the graphene support is still undamaged underneath the MoS$_2$ sheet (a copy of this micrograph is provided in supplementary material, with the contrast and brightness levels adjusted to reveal the graphene lattice more clearly). As with the bare MoS$_2$ experiment, we attempted to collect an EEL SI of the undamaged region highlighted by the blue rectangular box in the survey image in fig. 2f. The SI parameters were almost identical to the bare MoS$_2$ case (with slightly higher pixel time, pixel density and slightly wider SI spatial extent): the dose rate during the acquisition was therefore $5.5 \times 10^8$ e/Å$^2$.s, for a total accumulated dose of $9.2 \times 10^{10}$ e/Å$^2$. The 'after' image taken after collecting the spectra is presented in fig. 2h and shows again clear damage to the MoS$_2$ sheet, although perhaps not quite as severe as in the bare membrane case. The simultaneously-acquired HAADF image is shown in fig. 2g, revealing that the sheet was punctured again early through the SI acquisition, although at a higher total dose (after a larger number of pixels). While a few unit cells were mapped before the onset of damage, the resulting chemical maps (see supplementary material) would again not be exploitable for quantitative analysis. A single layer graphene support therefore enables longer imaging times for MoS$_2$, but it is not capable of preventing damage at the higher electron doses employed for chemical mapping. The same set of experiments was carried out on the same sample flipped over in the holder, so the beam would hit the graphene first and then the MoS$_2$, leading to identical results and conclusions.

(figure 2 here)

As a last step, single layer MoS$_2$ was encapsulated between single layer graphene sheets. The prepared sample stack was initially checked *via* electron diffraction to make sure that MoS$_2$ and graphene are present together as a stack (graphene/MoS$_2$/graphene) and all as single layers. As can be seen in the diffraction pattern in figure 3a, the stack consists of single layer MoS$_2$ (red dashed circle) [31] and two rotated graphene layers (blue dashed circle),[32] whose diffraction spots can be distinguished as they are separated by almost 23° and far less intense than the MoS$_2$ spots as mentioned in fig. 2. This time, we did not observe any damage to the MoS$_2$ structure during imaging even for the acquisition of images with longer exposure times and higher pixel densities (therefore at higher dose and dose rates) than in the previous two cases. Figure 3b and 3c are the first and the last frames from movie_3 (see supplementary information): they were taken 68 s apart after a total accumulated dose of $4.5 \times 10^8$ e/Å$^2$ and clearly no hole or defect is observed, supporting the claim that MoS$_2$ is undamaged under these conditions. However, an image taken immediately after recording the movie (exposed to a dose of $\sim 5.2 \times 10^6$ e/Å$^2$) showed some interesting contrast features as seen in fig. 3d. The area encircled by a blue dashed line in fig. 3d appears slightly darker than the surrounding area: we attribute this variation to a creation of a hole in one of the protective graphene sheets. Due to the double transfer process to fabricate this sample, the graphene membranes were covered with a slightly larger concentration of contaminants than the previous two samples, Si atoms and SiO$_2$ clusters in particular. These contaminants are known to participate as catalysts in a localized graphene etching mechanism in the experimental conditions employed here, resulting in this case in the formation of a hole in the graphene layer.[29,33] The encapsulated MoS$_2$ appears mostly unaffected, though. Moreover, the contribution of the graphene to the atomic contrast in the HAADF images is minimal, as could be expected from simple considerations based on the large difference in atomic weights of the elements involved (2x$Z_S$=32, $Z_{Mo}$=42 compared to $Z_C$=6) and the approximate $Z^2$ dependence of HAADF images. Some minor loss of sharpness and contrast is noticeable nonetheless. This effect can be easily quantified by drawing an intensity line profile across a Mo-2S 'dumbbell' and comparing the signal-to-background ratios for the Mo and 2S columns, defined as the ratio between the column peak intensity and the intensity in the middle of the hexagons (*i.e.* in a hole), having subtracted any dark current offset. Using figure 1a as a representative example of the contrast obtained in bare MoS$_2$ we determined signal-to-background ratios of 2.9 and 1.8 for Mo and 2S, respectively. When the MoS$_2$ sheet is encapsulated with graphene, such as on figure 3d, these signal-to-background values drop to 2.5 and 1.5 for Mo and 2S, respectively, in otherwise almost identical imaging conditions. This corresponds to a loss of contrast of approximately 15% between the bare and encapsulated cases. It is interesting to note that the graphene lattice is not clearly visible in encapsulated images (or indeed where only one layer of graphene is used as support). This is even clearer in images of an area where the MoS$_2$ sheet terminates, leaving only the graphene support visible (for more details, see Supplementary Material, figure S3). Both these experimental observation are borne out by multislice images simulations of single-layer MoS$_2$ and of a model of the encapsulated structure: see Supplementary Material. To make sure that graphene encapsulation of MoS$_2$ supports our basic idea of defect-free chemical mapping, an EEL SI was acquired in the blue rectangular box in the survey image in fig. 3e. As shown in fig. 3g, which was taken after the mapping, no damage was created in the mapped region, which is clearly apparent from the very sharp simultaneously acquired HAADF image of fig. 3f: both Mo and S sub-lattices are resolved throughout the SI acquisition. The apparent stability of the $MoS_2$ sheet even allowed to increase both the total dose and dose rate for the EELS acquisition by choosing more demanding parameters: 0.08 s dwell time per spectrum and higher pixel density. This resulted in a total electron dose of $1.7 \times 10^{11}$ e/Å$^2$, which is substantially higher than the one used for the SIs in figs. 1 and 2. Thus, the encapsulation of the single layer $MoS_2$ enabled us to investigate the sample without defect formation and, being able to employ high electron doses, we managed to obtain images and SIs with adequately high S/N ratios, to obtain accurate chemical information at the atomic scale (see supplementary information for the resulting Mo and S chemical maps). The underlying mechanisms responsible for such effective protection against beam damage are difficult to determine with certainty. The remarkable conduction properties of graphene (both thermal and electric) are certainly expected to contribute to a very effective dissipation of accumulated charge or heat under the beam. Graphene is therefore the ideal conductive 'coating' to help mitigate ionization, as suggested for instance by Egerton *et al.*[21] A full encapsulation will result in further advantages: the impermeability of graphene provides very effective protection against environmental effects such as chemical etching under the beam (due for instance to residual gases in the microscope column - unlikely in our case thanks to the near UHV conditions at the sample).[34] Should S or Mo atoms be displaced by knock-on (despite the low probability of such an event given the low beam energy), or ionised, the top and bottom carbon layers will also provide added stability to the structure and possibly prevent the displaced or ionised atoms from complete ejection. Finally, we also note that the close proximity between the graphene and $MoS_2$ layers may lead to the formation of interlayer bonds (and consequently a modification of the electronic structure of the encapsulated material). Although this suggestion is only speculative, such bond formation would be expected to favour electron transport between the graphene and $MoS_2$ layers and thus contribute to mitigating ionization.

(figure 3 here)

**Conclusions**

In summary, employing a 60 keV electron beam and near-UHV conditions ($<5 \times 10^{-9}$ torr at the sample) to reduce knock-on damage and to minimise ionization damage did not prevent any occurrence of severe damage to bare, free-standing $MoS_2$. Radiation damage was somewhat mitigated in images obtained after placing the MoS$_2$ layer on a single layer of graphene. However, the damage reduction was not sufficient at the high electron doses typically required for quantitative chemical mapping. Our results demonstrate that damage of single layer MoS$_2$ (whether arising from knock-on or ionization effects in the electron microscope), can be prevented altogether by encapsulation between graphene layers. The three-layer stack (graphene/MoS$_2$/graphene) allows the application of high electron doses for high resolution, defect free imaging and, importantly, for chemical analysis of MoS$_2$. We envisage this technique could also be employed for detailed studies of other beam sensitive materials, *e.g.*, molecules and nano- and bio- particles.

**Methods**

**Sample Preparation**

Both MoS$_2$ and graphene single layer flakes were prepared by mechanical exfoliation. For the first sample, following exfoliation, the single layer MoS$_2$ was directly transferred to a standard Quantifoil$^{TM}$ TEM grid *via* polymer based wet-transfer technique as samples need to be freely suspended for transmission electron microscopy measurements.[35] The second sample was fabricated by transferring by mask aligning techniques a single MoS$_2$ layer onto a single layer of graphene, which had been positioned (exfoliated) on Si/SiO$_2$ wafer. The two-layer stack was then transferred to a TEM grid in the same way as the single layer MoS$_2$. The third sample required a fabrication process consisting of two transfers: firstly, MoS$_2$ was transferred onto single layer graphene and secondly a further single layer graphene sheet was placed on top of the MoS$_2$ layer. This was followed by the removal of the Si/SiO$_2$ wafer substrate, thus releasing completely the three-layer stack for wet transfer onto a TEM grid. Each transfer during the sample fabrication was followed by a dip into acetone to remove protective polymer layers (PMMA). Once transferred to the TEM grid, the samples were dipped one final time in acetone and dried in a critical point dryer to avoid the surface tension damage to the flakes.

**Characterization**

Microscopy measurements were performed at the SuperSTEM Laboratory, on a Nion UltraSTEM100$^{TM}$ aberration-corrected dedicated scanning transmission electron microscope. The design of the column allows for clean high vacuum conditions at the sample (<5x10$^{-9}$ torr), reducing the probability of damage through chemical etching and preventing build-up of contamination which hinders high resolution observations. The Nion UltraSTEM has a cold field emission gun with a native energy spread of 0.35 eV and was operated at 60 keV primary beam energy. The beam was set up to a convergence semi-angle of 30 mrad with an estimated beam current of ~100 pA. Note that in a cold field emission instrument the probe current drops slightly with time until the tip is cleaned ('flashed'): all electron doses estimated here assume a freshly flashed tip and a current of 100 pA (the tip was systematically flashed shortly before all the acquisition of the data presented). Under these operating conditions, the estimated probe size is ~1.1 Å, providing the perfect tool for atom-by-atom chemical analysis;[14] these conditions are particularly adequate for MoS$_2$ as the distance between Mo and S atoms is 1.8 Å when the single layer slab is viewed along an (001) direction. These experimental conditions (scanning probe, low primary beam energy, high vacuum conditions) are significantly different from those in most other studies of $MoS_2$, which are typically performed with stationary and slightly higher energetic beams under poorer vacuum conditions. High Angle Annular Dark Field (HAADF) imaging was employed to produce atomically resolved images whose intensity is approximately proportional to the square of the average atomic number Z of the material under investigation. This chemically-sensitive 'Z-contrast' mode is ideally suited to directly identify the nature of individual atoms.[14] HAADF imaging is complemented by further chemical fingerprinting through Electron Energy Loss Spectroscopy (EELS).

**Conflict of Interest**

The authors declare no competing financial interest.

**Supporting Information Available**

This additional material consists of Movies (.avi) showing the dynamic behaviour of the samples under the electron beam. Further images, electron energy loss chemical maps and multislice simulations are also provided in a separate .pdf document. This material is available free of charge *via* the Internet at http://pubs.acs.org.

**Acknowledgements**

This work was supported by the EPSRC (UK). The SuperSTEM Laboratory is the EPSRC National Facility for Aberration-Corrected STEM.

**Figure Captions**

**Figure 1.** Atomic resolution HAADF images (raw data) of pristine single layer $MoS_2$. a) before and b) after consecutive scans, showing vacancy formation; c) first and d) last frame of movie_1 showing expansion of the damaged region; e) before, f) during g) after acquisition of an EEL spectrum image (SI) (f) is the HAADF image of the SI area (blue frame in (e)), taken simultaneously with the SI). The electron dose was $2.6 \times 10^{10}$ e/Å$^2$.

**Figure 2.** Atomic resolution Z-contrast images (raw data) of single layer $MoS_2$ on graphene. a) diffraction pattern showing spots from both $MoS_2$ (red dashed circle) and graphene (blue dashed circle); image b) before and c) after consecutive scans, showing no vacancy formation; d) the first and e) the last frame of movie_2 showing hole formation; image f) before, g) during and h) after acquisition of the EEL SI, showing damage in $MoS_2$ at high electron dose ($9.2 \times 10^{10}$ e/Å$^2$).

**Figure 3.** Atomic resolution Z-contrast images (raw data) of single layer $MoS_2$ encapsulated by graphene layers. a) diffraction pattern showing spots from both $MoS_2$ (red dashed circle) and two graphene layers rotated by 23° with respect to each other (blue dashed circle); b) first and c) last frame of movie_3 showing no vacancy formation; d) after the movie showing damage in the graphene layer; e) before, f) during and g) after the EEL SI acquisition, showing no damage in the $MoS_2$ even at the high electron dose of $1.7 \times 10^{11}$ e/Å$^2$.

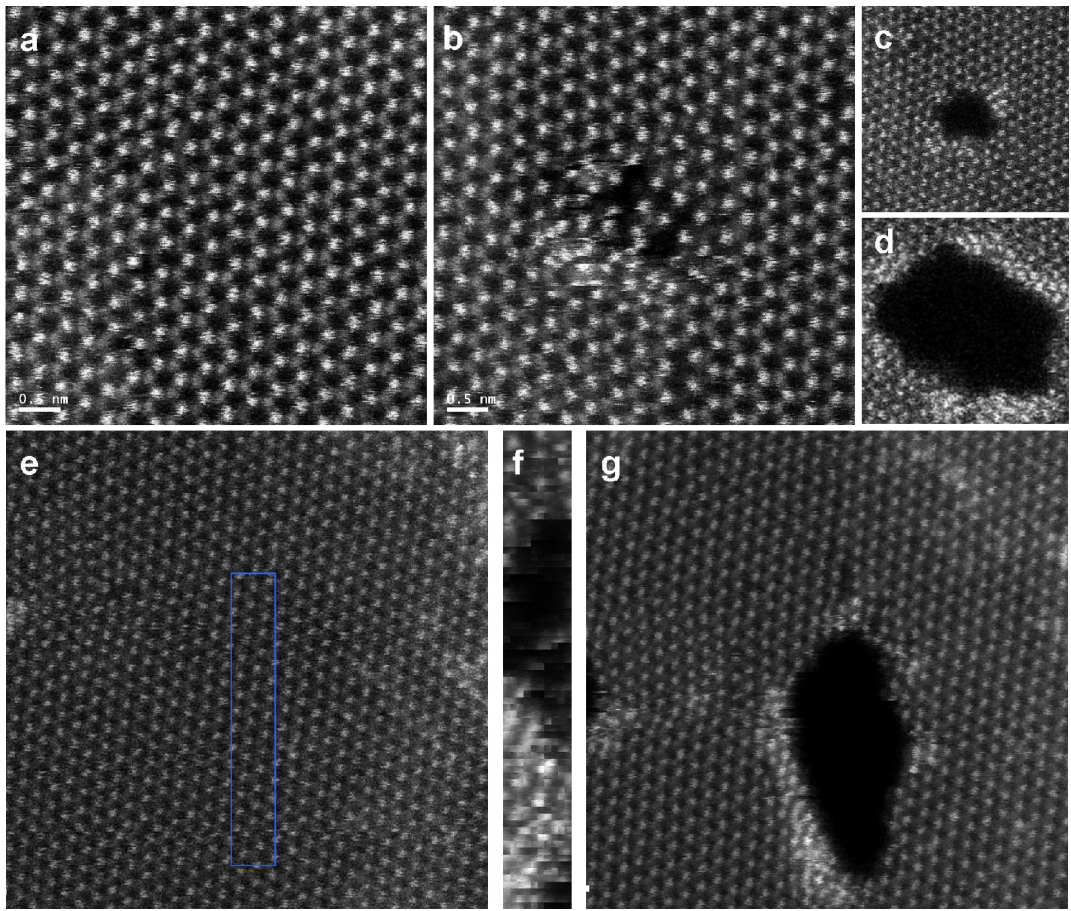

**Figure 1.**

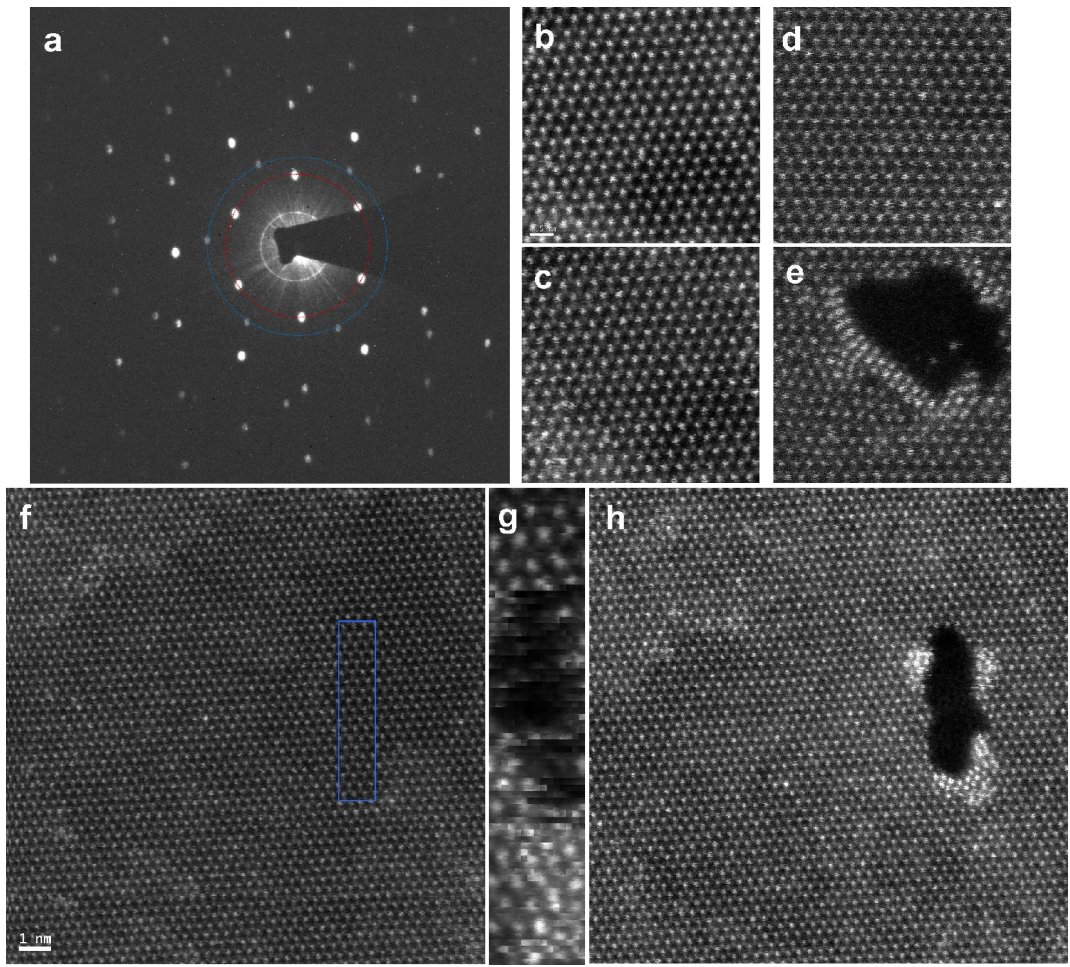

**Figure 2.**

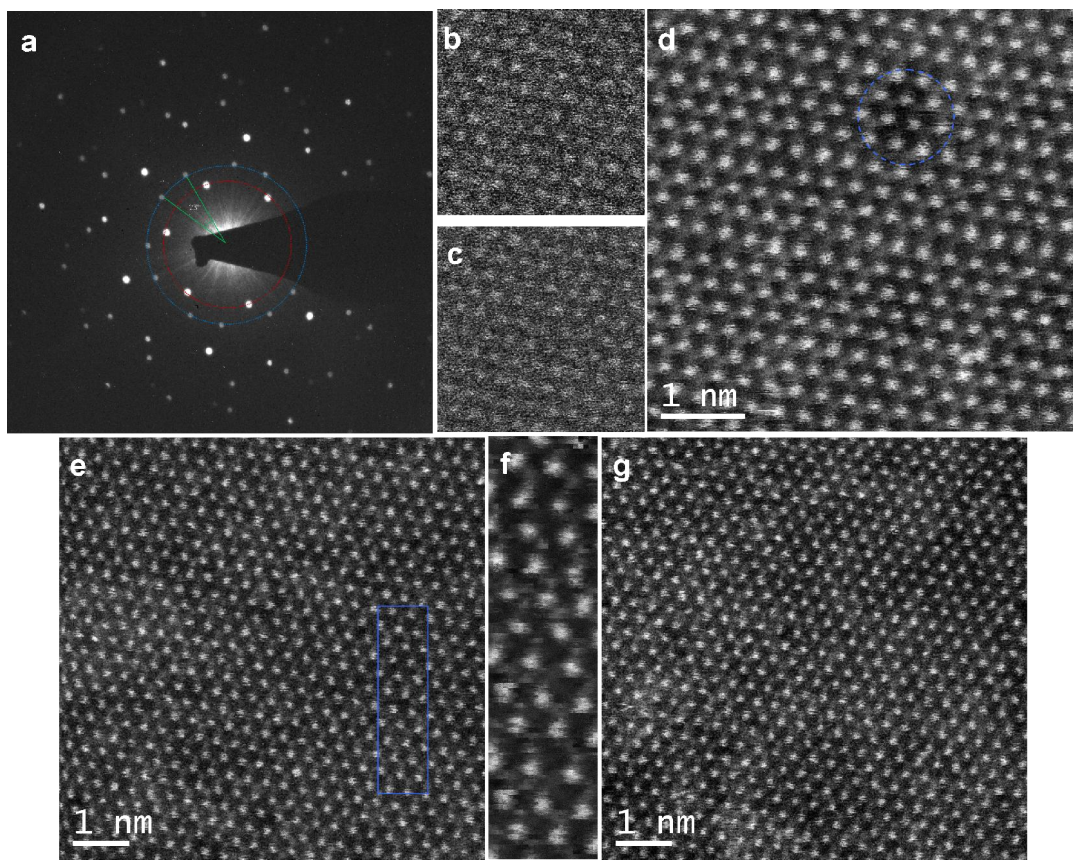

**Figure 3.**

**TOC**

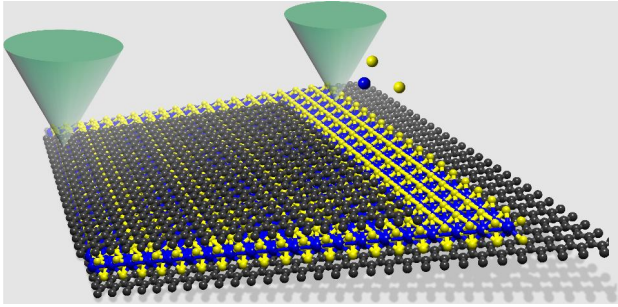

Encapsulating a sheet of MoS2 within two single layers of graphene helps control radiation damage sustained when observing this material in a scanning transmission electron microscope.